\documentclass[aps, prb, twocolumn, showpacs, superscriptaddress]{revtex4-1}

\usepackage{amsmath,amssymb}
\usepackage{graphicx}% Include figure files
\usepackage{dcolumn}% Align table columns on decimal point
\usepackage{bm}% bold math
\begin{document}

\title {Correlated metallic state in honeycomb lattice: Orthogonal Dirac semimetal}
\author{Yin Zhong}
\email{zhongy05@hotmail.com}
\affiliation{Center for Interdisciplinary Studies $\&$ Key Laboratory for
Magnetism and Magnetic Materials of the MoE, Lanzhou University, Lanzhou 730000, China}
\author{Ke Liu}
\affiliation{Institute of Theoretical Physics, Lanzhou University, Lanzhou 730000, China}
\author{Yong-Qiang Wang}
\affiliation{Institute of Theoretical Physics, Lanzhou University, Lanzhou 730000, China}
\author{Hong-Gang Luo}
\email{luohg@lzu.edu.cn}
\affiliation{Center for Interdisciplinary Studies $\&$ Key Laboratory for
Magnetism and Magnetic Materials of the MoE, Lanzhou University, Lanzhou 730000, China}
\affiliation{Beijing Computational Science Research Center, Beijing 100084, China}

\date{\today}

\begin{abstract}
A novel gapped metallic state coined orthogonal Dirac semimetal is proposed in the honeycomb lattice in terms of $Z_{2}$ slave-spin representation of Hubbard model. This state corresponds to the disordered phase of slave-spin and has the same thermaldynamical and transport properties as usual Dirac semimetal but its singe-particle excitation is gapped and has nontrivial topological order due to the $Z_{2}$ gauge structure. The quantum phase transition from this orthogonal Dirac semimetal to usual Dirac semimetal is described by a mean-field decoupling with complementary fluctuation analysis and its criticality falls into the universality class of 2+1D Ising model while a large anomalous dimension for the physical electron is found at quantum critical point (QCP), which could be considered as a fingerprint of our fractionalized theory when compared to other non-fractionalized approaches. As byproducts, a path integral formalism for the $Z_{2}$ slave-spin representation of Hubbard model is constructed and possible relations to other approaches and the sublattice pairing states, which has been argued to be a promising candidate for gapped spin liquid state found in the numerical simulation, are briefly discussed. Additionally, when spin-orbit coupling is considered, the instability of orthogonal Dirac semimetal to the fractionalized quantum spin Hall insulator (fractionalized topological insulator) is also expected. We hope the present work may be helpful for future studies in $Z_{2}$ slave-spin theory and related non-Fermi liquid phases in honeycomb lattice.
\end{abstract}

\maketitle

\section{Introduction} \label{intr}
It is still a challenge to understand exotic quantum phases and its corresponding quantum criticality beyond the conventional Ginzburg-Landau-Wilson paradigm. \cite{Sachdev2011,Sachdev2003,Rosch,Sachdev2008,Powell}
One of ideas \cite{Sachdev2011,Georges1996,Schollwock,Wen2004,Wen,Kitaev,Levin,Metzner} to attack this challenging problem is to fractionalize the electrons in the model Hamiltonian into more elementary collective excitations, e.g., quasiparticles like spinon, holon, and so on, in terms of slave-particle approaches near the putative quantum critical points due to wild quantum fluctuations. \cite{Wen,Senthil2003,Senthil2004,Senthil3,Senthil4,Florens,Lee2005,Pepin2005,Kim2006,Senthil2008,Kim2010,Senthil2010}

Recently, a remarkable determinantal quantum Monte Carlo simulation has been performed in the honeycomb lattice at half-filling.\cite{Meng} This study claimed that there exists a short-range resonating valence-bond liquid with a spin gap between the Dirac semimetal and antiferromagnetic insulating phases. \cite{Sorella1992,Herbut2006,Kotov2010}(A similar gapped spin liquid phase has also been found in the Kagome Heisenberg model by density-matrix-renormalization-group.\cite{Yan} However, it is noted that a very recent quantum Monte Carlo simulation, which extend the previous study to much larger clusters, finds no evidence of a spin liquid region found in Ref.[\onlinecite{Meng}].\cite{Sorella2012}) Immediately, the nature of this possible spin liquid state has been explored by using slave-particle framework with U(1) and/or SU(2) gauge structure and was proposed to be a $Z_{2}$-like spin liquid, \cite{Wang2010,Tran2011,Lu2011} which was carefully inspected by later numerical simulations in $J_{1}$-$J_{2}$ Heisenberg model. \cite{Clark,Mezzacapo2012}(Note, however, simple reduction of the physics of the Hubbard model to that of an effective spin-1/2 system, such as the $J_{1}$-$J_{2}$, might not be achievable in light of Ref. [\onlinecite{Mezzacapo2012}].)\cite{Mezzacapor}

Instead of studying the elusive insulating spin liquid phase found in the numerical simulation,\cite{Meng} it is also interesting to see whether a possible fractionalized metallic state can exist in the honeycomb lattice. This is stimulated by a recent paper of Nandkishore, Metlitski and Senthil, \cite{Nandkishore} where they reinspected the $Z_{2}$ slave-spin representation of single-band Hubbard model and pointed out that the correct disordered state of slave spins is not a putative Mott insulator but an exotic fractionalized metallic state called orthogonal metal. \cite{deMedici,Hassan,Ruegg,Yu} This state is a compressible metal, which has the same thermodynamics and transport as the usual Landau Fermi liquid, but its electronic spectral function has a gap, which results in a simplest non-Fermi liquid. Therefore, there is an orthogonal metal-Fermi liquid transition instead of Mott transition in the slave spin representation of the single-band Hubbard model. Furthermore, for the multi-orbital models like the Anderson lattice model, an orbital-selective orthogonal metal transition has been found by the $Z_{2}$ slave-spin approach \cite{Zhong2012} and is argued to be an alternative Kondo breakdown mechanism for certain heavy fermion compounds. \cite{Senthil2003,Senthil2004,Paul,Pepin,Pepin2008,Paul2008,Kim2010,Senthil2010}

In the present paper we try to uncover a possible fractionalized metallic state in terms of $Z_{2}$ slave-spin representation of the Hubbard model in the honeycomb lattice at half-filling following the same methology of Nandkishore, Metlitski and Senthil. \cite{Nandkishore} It is found that while the ordered state of slave-spin is identified as the usual Dirac semimetal, an exotic gapped semimetal named orthogonal Dirac semimetal survives when the slave spin becomes disordered. The orthogonal Dirac semimetal has the same thermodynamics and transport as the usual Dirac semimetal but with the nontrivial topological order due to the local $Z_{2}$ gauge structure.

Furthermore, we analyze in detail the corresponding quantum phase transition (QPT) between these two metallic states at mean-field level and find that it falls into the universality class of 2+1D Ising model while a large anomalous dimension for the physical electron is found at quantum critical point(QCP). This result could be considered as a fingerprint of our fractionalized treatment when compared to other non-fractionalized approaches. Additionally, the mean-field analysis is stable to weak fluctuations since vanishing density of states of the Dirac semimetal suppresses the effect of Landau damping even under the condition without long-range Coulomb interaction. This is in contrast to the original case of orthogonal metal where the mean-field treatment of quantum criticality is only valid in the presence of the long-range Coulomb interaction. As byproducts, a path integral formalism for the $Z_{2}$ slave-spin representation of Hubbard model is constructed and possible relations to the exotic spin liquid named sublattice pairing states, fractionalized quantum spin Hall insulator, effective Gross-Neveu theory, the dual approach of interacting fermions and the gauge/gravity duality are briefly discussed. Moreover, we expect the orthogonal Dirac semimetal may be realized in future experiments of ultracold atoms in the honeycomb optical lattices \cite{Delgado} and surface states of three dimensional topological insulators since they naturally support Dirac fermions as their low-lying excitations.\cite{Bloch,Goldman,Bermudez,Neto,Hasan2010,Qi2011} In addition, it may be interesting to generalize the present analysis of orthogonal Dirac semimetal to the bilayer graphene-a semimetal with a non-vanishing density of state, which is known to have a strongly correlated ground state.\cite{Nandkishore2010,Nandkishore2012}

The remainder of this paper is organized as follows. In Sec. \ref{sec2}, we introduce the Hubbard model in the honeycomb lattice and reformulated it in terms of the $Z_{2}$ slave-spin representation. Moreover, a path integral formalism for this slave-spin representation of the Hubbard model is also constructed in this section. Then, a mean-field decoupling is used and two resulting mean-field states are analyzed in Sec. \ref{sec3}. One state is the expected orthogonal Dirac semimetal and the other is the usual Dirac semimetal. In Sec. \ref{sec4}, the QPT between these two states are discussed with emphasizing on the scaling behaviors and the stability of mean-field treatment of quantum criticality is studied. Some relations to other approaches and instability of the orthogonal Dirac semimetal are briefly discussed In Sec. \ref{sec5}. Finally, Sec. \ref{sec6} is devoted to a conclusion which summarizes our main findings.

\section{$Z_{2}$ slave spin representation and the Hubbard model in the honeycomb lattice} \label{sec2}
The model we used is the Hubbard model in the honeycomb lattice at half-filling, \cite{Herbut2006,Ruegg}
\begin{eqnarray}
&&H=-t\sum_{\langle ij\rangle\sigma}(c_{i\sigma}^{\dag}c_{j\sigma}+h.c.)+\frac{U}{2}\sum_{i}(n_{i}-1)^{2}, \label{eq1}
\end{eqnarray}
where $n_{i}=\sum_{\sigma}c_{i\sigma}^{\dag}c_{i\sigma}$, $U$ is the onsite Coulomb energy between electrons on the same site and $t$ is the hopping energy between nearest-neighbor sites. Since we are interested in the case of half-filling, the chemical potential has been set to zero. This model has been studied by many authors and is believed to exhibit several distinct phases depending on the ratio of $U/t$. (For a review, see Ref. [\onlinecite{Kotov2010}].) For small $U/t$, the usual Dirac semimetal appears with nearly free relativistic Dirac fermions being the low energy excitation. In contrast, an antiferromagnetic Mott insulator survives for large $U/t$ and spin rotation symmetry breaks spontaneously.\cite{Sorella1992} Besides, some exotic spin liquid Mott insulating states with a charge gap, e.g. algebraic spin liquid (ASL) and $Z_{2}$ spin liquids, may exist in the intermediate coupling as argued by slave-particle techniques with the projective symmetry group (PSG) analysis and inspected by sophisticated numerical simulations. \cite{Hermele2007,Meng,Wang2010,Tran2011,Lu2011,Clark}

Here, the problem we are interested in is whether there exists an alternative metallic state beside the trivial Dirac semimetal. It seems that such a metallic state is likely due to the fractionalizing scheme proposed by Nandkishore, Metlitski and Senthil, \cite{Nandkishore} in which they found a $Z_{2}$ fractionalized metal named as the orthogonal metal in the $Z_{2}$ slave-spin representation of the Hubbard model. The orthogonal metal was incorrectly identified as a non-magnetic Mott insulator by previous studies because the slave-spin was considered to carry charge of the physical electron. As a matter of fact, as also pointed out by Nandkishore, Metlitski and Senthil, \cite{Nandkishore} the slave-spin does not carry any quantum numbers of the physical electron but only has the $Z_{2}$ gauge charge since it is bound by the gapped $Z_{2}$ gauge field (a brief explanation of this point is given by Appendix A) Thus, all quantum numbers of the physical electron have to be taken by the slave-fermion, which also carries a $Z_{2}$ gauge charge to ensure the gauge invariance of the physical electron composed by a slave-spin and a slave-fermion.

Therefore, it is expected that a fractionalized metallic state similar to the orthogonal metal may exist in the honeycomb lattice when using $Z_{2}$ slave-spin representation. To this point, a careful reader may wonder what the difference is between our study and the case of Nandkishore, Metlitski and Senthil \cite{Nandkishore} since the same starting points have been made, namely, the Hubbard model and the $Z_{2}$ slave-spin representation. We should emphasize that in the paper of Nandkishore, Metlitski and Senthil, \cite{Nandkishore} a square lattice is kept in mind implicitly and the reference state is the usual Landau Fermi liquid with a large Fermi surface since a Fermi liquid is indeed a good starting point for a small $U/t$ in the square lattice. However, in the present work, we will discuss the Hubbard in the honeycomb lattice at half-filling and instead of the Fermi liquid, a Dirac semimetal is our reference state, which has relativistic Dirac fermion excitation near disconnected Dirac points. Thus, it will lead to a different fractionalized metallic state from the orthogonal metal. This will be seen in the following discussion. Now, we turn to the $Z_{2}$ slave-spin framework for a honeycomb lattice at half-filling.

\subsection{$Z_{2}$ slave-spin representation of the Hubbard model}
In the treatment of $Z_{2}$ slave-spin approach, the physical electron $c_{\sigma}$ is fractionalized into a new slave fermion $f_{\sigma}$ and a slave spin $\tau^{x}$ as \cite{deMedici,Ruegg}
\begin{equation}
c_{i\sigma}=f_{i\sigma}\tau_{i}^{x}\label{eq2}
\end{equation}
with a constraint $\tau_{i}^{z}+1=2(n_{i}-1)^{2}$ enforced in every site. Under this representation, the original Hamiltonian can be rewritten as
\begin{eqnarray}
&&H=-t\sum_{\langle ij\rangle \sigma}(\tau_{i}^{x}\tau_{j}^{x}f_{i\sigma}^{\dag}f_{j\sigma}+h.c.)+\frac{U}{4}\sum_{i}(\tau_{i}^{z}+1)\label{eq3}
\end{eqnarray}
where $n_{i}=n_{i}^{f}=\sum_{\sigma}f_{i\sigma}^{\dag}f_{i\sigma}$. Obviously, a $Z_{2}$ local gauge symmetry is left in this representation (both slave-fermions and slave spins carrying the $Z_{2}$
gauge charge) and the corresponding low-energy effective theory should respect this. The mentioned gauge structure can be seen if $f_{i\sigma}^{(\dag)}\rightarrow \epsilon_{i}f_{i\sigma}^{(\dag)}$ and $\tau_{i}^{x}\rightarrow\epsilon_{i}\tau_{i}^{x}$ with $\epsilon_{i}=\pm1$ while the whole Hamiltonian $H$ is invariant under this $Z_{2}$ gauge transformation. Moreover, as argued by Nandkishore, Metlitski and Senthil, \cite{Nandkishore} the slave-fermion $f_{\sigma}$ carries both charge and spin of the physical electron $c_{\sigma}$ while slave spin $\tau^{x}$ encodes the remanent quantum coherence of physical electrons, which has also been mentioned on the issue of fractionalized quantum spin Hall insulator. \cite{Ruegg2012}

\subsection{Path integral formalism for the $Z_{2}$ slave-spin representation of the Hubbard model}
Before turning into discuss the possible alternative metallic state, we present the construction of a path integral formalism for the $Z_{2}$ slave-spin approach of the Hubbard model in this subsection. To this aim, we follow the approach of Ref. [\onlinecite{Senthil2000}] where the general $Z_{2}$ gauge theory is constructed in an extended Hubbard model.

The construction of the path integral is to calculate the partition function $Z=\text{Tr}(e^{-\beta \hat{H}}\hat{P})$ where
$\hat{P}$ is the projective operator to exclude unphysical states introduced by $Z_{2}$ slave-spin representation.
Here we use
\begin{equation}
\hat{P}=\prod_{i}\frac{1}{2}(1+(-1)^{\frac{1}{2}[\tau_{i}^{z}+1-2(n_{i}^{f}-1)^{2}]}).\label{eq4}
\end{equation}
This choice has the advantage to meet the mean-field theory of $Z_{2}$ slave-spin approach. Obviously, one can employ another equivalent projective operator\cite{Ruegg2012}
\begin{equation}
\hat{P}=\prod_{i}\frac{1}{2}(1+(-1)^{\frac{1}{2}[\tau_{i}^{z}-1+2n_{i}^{f}]}).\label{eq5}
\end{equation}
We will use the first definition of $\hat{P}$ in the following discussion.
Follow Ref. [\onlinecite{Senthil2000}], the projective operator can be reformulated by introducing auxiliary Ising field $\sigma_{i}=\pm1$
\begin{equation}
\hat{P}=\prod_{i}\frac{1}{2}\sum_{\sigma_{i}=\pm1}e^{i\frac{\pi}{4}(\sigma_{i}-1)[\tau_{i}^{z}+1-2(n_{i}^{f})^{2}]}.\label{eq6}
\end{equation}
Since $[\hat{P},H]=0$, one can define an effective Hamiltonian $H_{eff}$ as
\begin{eqnarray}
H_{eff}=H+\sum_{i}i\frac{\pi}{4}(1-\sigma_{i})[\tau_{i}^{z}+1-2(n_{i}^{f})^{2}].\label{eq7}
\end{eqnarray}
Then using the same method in the treatment of quantum Ising model (see Appendix B) and standard coherent state representation of fermions, one obtains the following path integral formalism of $Z_{2}$ slave-spin representation of Hubbard model\cite{Rueggpr}
\begin{equation}
Z=\prod_{ni}\int d\bar{f}_{i}(n)df_{i}(n)d\varphi_{i}(n)\delta(\varphi^{2}_{i}-1)d\sigma_{i}(n)\delta(\sigma^{2}_{i}-1) e^{-S}\label{eq8}
\end{equation}
and
\begin{eqnarray}
&&S=\sum_{n,i}\bar{f}_{i\sigma}(n)(f_{i\sigma}(n)-f_{i\sigma}(n-1)) \nonumber\\
&&\hspace{1cm} +\sum_{n,i}\varphi_{i}(n)\varphi_{i}(n+1)a(n) \nonumber\\
&&\hspace{1cm} -\epsilon t\sum_{n\langle ij\rangle}(\varphi_{i}(n)\varphi_{j}(n)\bar{f}_{i\sigma}(n)f_{j\sigma}(n)+c.c.) \nonumber\\
&&\hspace{1cm} +\epsilon\sum_{n,i}i\frac{\pi}{4}(1-\sigma_{i}(n))[1-2(n_{i}^{f})^{2}],\label{eq9}
\end{eqnarray}
where we have used $\tau_{i}^{x}|\varphi\rangle=\varphi_{i}|\varphi\rangle$ with $\varphi=\pm1$ and $\tau_{i}^{z}|\varphi\rangle
=|\varphi_{1}\rangle|\varphi_{2}\rangle|\varphi_{3}\rangle\cdot\cdot\cdot|-\varphi_{i}\rangle\cdot\cdot\cdot|\varphi_{N}\rangle$ to avoid confusion with auxiliary Ising field $\sigma_{i}$ with $a(n)=-\frac{1}{2}\ln\epsilon(\frac{U}{4}+i\frac{\pi}{4\beta}(1-\sigma_{i}(n)))$. The above action is our main result in this subsection and further approximations have to be made in order to gain some physical insights. A popular approximation is to decouple the interaction term between slave-spin $\varphi$ and slave-fermion $f_{\sigma}$ at the mean field level and then reintroduce phase fluctuations (in fact, a $Z_{2}$ gauge field due to the gauge structure of $H$). \cite{Senthil2000,Ruegg2012} Details of this treatment has not been reported until now, however, we hope our formalism constructed here may be useful in this direction.

\section{mean-field theory and orthogonal Dirac semimetal} \label{sec3}
Undoubtedly, it is a formidable task to treat exactly the Hamiltonian of the Hubbard model in the slave-spin representation given by Eq. (\ref{eq3}) and its path integral formalism given by Eq. (\ref{eq8}). Thus we only consider a mean-field treatment here and discuss the stability of the mean-field analysis in the next section.

It is straightforward to derive a mean-field Hamiltonian as follows \cite{Ruegg}
\begin{eqnarray}
&&H_{f}=-\sum_{\langle ij\rangle\sigma}(\tilde{t}_{ij}f_{i\sigma}^{\dag}f_{j\sigma}+h.c.)-2\sum_{i}\lambda_{i}(n_{i}^{f}-1)^{2},\label{eq10} \\
&&H_{I}=-\sum_{\langle ij\rangle\sigma}(J_{ij}\tau_{i}^{x}\tau_{j}^{x}+h.c.)+\sum_{i}\left(\lambda_{i}+\frac{U}{4}\right)\tau_{i}^{z},\label{eq11}
\end{eqnarray}
where the Lagrange multiplier $\lambda_{i}$ has been introduced to fulfill the constraint on average,
$\tilde{t}_{ij}=t\langle \tau_{i}^{x}\tau_{j}^{x}\rangle$, $J_{ij}=t\sum_{\sigma}\langle f_{i\sigma}^{\dag}f_{j\sigma}\rangle$ and an extra self-consistent equation appears as   $\langle\tau_{i}^{z}\rangle+1=2\langle(n_{i}-1)^{2}\rangle$ due to the constraint. The decoupled Hamiltonian $H_{I}$ is an extended quantum Ising model in transverse field and $H_{f}$ describes $f$ fermions in the honeycomb lattice. Here, at the mean-field level, a further simplification can be made by setting all the Lagrange multiplier $\lambda_{i}$ to be zero, provided only non-magnetic solutions are involved and a half-filling case is considered. \cite{Ruegg} This means the constraint is not violated seriously when magnetic order is absent. Therefore, we will drop the constraint term in the mean-field Hamiltonian Eqs. (\ref{eq10}) and (\ref{eq11}) hereafter.

\subsection{Quantum Ising model in the honeycomb lattice}
Let us first focus on the quantum Ising model given by Eq.(\ref{eq11}). It is well known that the standard transverse field Ising model in one spatial dimension can be exactly solved by Jordan-Wigner transformation and it has two phases with the critical exponents being the same as two-dimensional classical Ising model. \cite{Sachdev2011} Beyond one spatial dimension, to our knowledge, no exact solutions exist for the quantum Ising model until now. However, generically, one may define $\langle \tau^{x}\rangle $ as a useful order parameter and there are at least two phases in two space dimensions or beyond. (It is just this case in the study of single-band Hubbard model in terms of some mean-field approximations and the Schwinger bosons theory. \cite{deMedici,Ruegg,Nandkishore}) One is a magnetic state with $\langle \tau^{x}\rangle \neq0$ while the other is described by a vanished $\langle \tau^{x}\rangle$ and is a disordered state with an excitation gap. Moreover, there exists a QCP between these two distinct phases, whose critical properties could be described by a quantum $\varphi^{4}$ theory,

In the case of the extended quantum Ising model in the honeycomb lattice, R\"{u}egg and Fiete \cite{Ruegg2012} found that there exist a ferromagnetic phase $\langle \tau^{x}\rangle \neq0$ and a paramagnetic phase $\langle \tau^{x}\rangle=0$ by using a 4-site cluster-mean-field approximation for the mean-field Hamiltonian. However, the paramagnetic phase is interpreted as the valence-bond solid (VBS), which turns out to break both lattice rotation and translation symmetry. Thus, if so, the putative quantum critical point between these two states will be unstable since the ferromagnetic phase and the valence-bond solid breaks entirely different symmetries and a first-order transition is expected generally according to the Ginzburg-Landau-Wilson paradigm. (A deconfined criticality \cite{Senthil3} cannot be excluded in principle but we have not seen any signals for this consideration.)

However, we note a different slave-spin treatment, which combines with a Schwinger boson analysis instead of a cluster-mean-field approximation, favors paramagnetic phases without broken lattice rotation and translation symmetry if the onsite $U$ is not sufficiently large.\cite{Vaezi2011} Furthermore, a slave-rotor approach has also been used to study the strong coupling behaviors of the Hubbard model in the honeycomb lattice. \cite{Lee2005,Pesin,Rachel} In these papers, the insulating state of quantum XY model, which corresponds to the paramagnetic phase of slave-spin representation, is found to preserve all physical symmetries. Thus, we expect a paramagnetic phase without any broken symmetries in the effective quantum Ising model in the honeycomb lattice with the belief that the slave rotor and slave-spin approach are equivalent though different interpretations are involved. \cite{Florens,Lee2005,deMedici,Hassan,Ruegg,Yu,Nandkishore}

Therefore, we may simply assume that the quantum Ising model in the honeycomb lattice has a ferromagnetic ordered phase ($\langle \tau^{x}\rangle \neq0$) and a disordered paramagnetic phase without any broken symmetries ($\langle \tau^{x}\rangle =0$) with a QCP between them. Moreover, an effective theory for the quantum Ising model in the honeycomb lattice can be derived by using its path integral formalism (see Appendix B for details)
\begin{eqnarray}
&&Z=\sum_{\{\varphi\}=\pm1}\prod_{n=1}^{N}e^{\epsilon \sum_{<ij>}J_{ij}(\varphi_{Ai}(n)\varphi_{Bj}(n)+\varphi_{Bi}(n)\varphi_{Aj}(n))}\nonumber\\
&&\hspace{1cm}\times e^{\sum_{i}a(\varphi_{Ai}(n)\varphi_{Ai}(n+1)+\varphi_{Bi}(n)\varphi_{Bi}(n+1))}, \label{eq12}
\end{eqnarray}
where $\varphi_{M}(M=A,B)$ is the slave-spin $\tau_{x}$ in A or B sublattice since the honeycomb lattice is a bipartite system and $a=-\frac{1}{2}\ln(\epsilon\frac{U}{4})>0$.

Then expanding the above action and working in continuum limit, one finds an effective theory as
\begin{equation}
Z=\int D\phi e^{-\int d\tau d^{d}x [\partial_{\tau}\varphi)^{2}+c^{2}(\nabla\varphi)^{2}+r\varphi^{2}+u\varphi^{4}]}, \label{eq13}
\end{equation}
where $c,r,u$ are effective parameters depending on microscopic details and $\varphi_{A}(x,\tau)\sim\varphi_{B}(x,\tau)\sim\varphi(x,\tau)$ since only the modes near zero momentum are involved in low-energy limit. Therefore, the quantum Ising model in the honeycomb lattice can also be described in terms of a $\varphi^{4}$ theory in spite of its bipartite feature, which may simplify corresponding calculations dramatically.

\subsection{Hamiltonian of slave-fermion in the honeycomb lattice}
Next, we treat the mean-field Hamiltonian $H_{f}$. It is noted that $H_{f}$ is a free Hamiltonian when dropping the contribution from the constraint. Then the resulting Hamiltonian describes a Dirac semimetal which has the following formalism in the low-energy limit \cite{Kotov2010}

\begin{equation}
S_{f}=\int d\tau\int d^{2}x \sum_{\sigma}\bar{\psi}_{\sigma}\gamma_{\mu}\partial_{\mu}\psi_{\sigma}, \label{eq14}
\end{equation}
where $\gamma_{0}=I_{2}\bigotimes\sigma_{z}$,$\gamma_{1}=\sigma_{z}\bigotimes\sigma_{y}$,$\gamma_{2}=I_{2}\bigotimes\sigma_{x}$
with $I_{2}$ the $2\times2$ identity matrix, $\sigma_{x},\sigma_{y},\sigma_{z}$ being the standard Pauli matrix. The Dirac fermion $\psi_{\sigma}$ is defined as $\psi_{\sigma}=[f_{A\sigma}^{1},f_{B\sigma}^{1},f_{A\sigma}^{2},f_{B\sigma}^{2}]^{T}$ with the transpose $T$ and we also have $\bar{\psi}_{\sigma}=\psi_{\sigma}^{\dag}\gamma_{0}$. $f_{M\sigma}^{\alpha}$ ($\alpha=1,2$) is the fermion in $M=A,B$ sublattices near two nonequivalent Dirac points located at $\vec{K}=-\vec{K}'=(1,1/\sqrt{3})(2\pi/\sqrt{3})$. It is clear that this free Dirac semimetal fixed point is stable for any short-range interactions, provided the coupling parameter is not large. \cite{Herbut2006}(This stability of free Dirac semimetal fixed point can be easily seen from a simple RG argument and it does not rely on the existence of long ranged Coulomb interactions in Ref.  [\onlinecite{Herbut2006}].) Thus, the free Dirac semimetal fixed point with disconnected Dirac points can serve as a good starting point in the honeycomb lattice at half-filling just like the usual Fermi liquid with large Fermi surface. In addition, we should emphasize it is the slave-fermion $f_{\sigma}$ that forms a Dirac semimetal but not the physical electrons $c_{\sigma}$ since in the slave-spin approach, the physical electron is a composite particle of slave-spin and slave-fermion ($c_{\sigma}=\tau^{x}f_{\sigma}$).

Because usually it is more interesting to study the instability of Dirac semimetal of physical electrons to other states via quantum phase transitions, we here assume $f$ fermions form a Dirac semimetal whatever the phases of slave spins are, and have a sharply defined Dirac quasiparticle in the remaining parts of the present paper. (A single-site mean-field solution of Eqs. [\ref{eq10}] and [\ref{eq11}] without the Lagrange multiplier can be found in Appendix D. There we find that the f-fermions form a Dirac semimetal when slave-spin orders and those fermions decouple from each other when slave-spin is disordered. This decoupling is obviously the artifact of the single-site approximation. If many-site effect is introduced, we expect that the f-fermions form the Dirac semimetal with renormalized parameters as what has been done in the paper of Nandkishore, Metlitski and Senthil.\cite{Nandkishore} Even though the f-fermions do not form the Dirac semimetal when slave-spin is disorder, the orthogonal Dirac semimetal may still be considered as a useful intermediate temperature state before other long-ranged ordered phases start to form at low temperature.)

\subsection{Nature of physical electrons in the honeycomb lattice and orthogonal Dirac semimetal}
To gain some features of the physical electrons in the $Z_{2}$ slave-spin representation for the honeycomb lattice, it is helpful to inspect the behavior of the quasiparticle, particularly, its single-particle Green's function or equivalently its spectral function.

In the case of $\langle\tau^{x}\rangle\neq0$ (ordered state of the corresponding quantum Ising model),
it is clear that $c_{i\sigma}\simeq\langle\tau^{x}\rangle f_{i\sigma}$, which means the slave-fermion corresponds to the physical electron when the slave-spin condensates. Therefore, the physical electron excitation is a
Dirac quasiparticle since slave-fermion forms Dirac semimetal
\begin{equation}
G(k)=Z\frac{i\gamma_{\mu}k_{\mu}}{k^{2}}\label{eq15}
\end{equation}
where the quasiparticle spectral weight is defined as $Z=\langle \tau^{x}\rangle ^{2}$ and $k^{2}=\vec{k}^{2}+\omega^{2}$ with $\omega$ being the imaginary frequency. Obviously, the obtained state with condensed slave-spins is just the usual Dirac semimetal since no fractionalized excitation will appear in physical observable. In the language of the gauge theory, the hidden $Z_{2}$ gauge field in the $Z_{2}$ slave-spin representation is confined by the Higgs mechanism when the slave-spin condensates, thus only gauge singlet of fractionalized particles ($f$ or $\tau^{x}$) are allowed in physical excitations due to the confined potential generated by the $Z_{2}$ gauge field.\cite{Senthil2000,Ruegg2012} One can refer to Appendix C for details.

In contrast, for a vanished $\langle \tau^{x}\rangle$ (disordered state of the slave spin) and in the low energy limit, the quantum Ising model is described by an effective $\varphi^{4}$ theory given by Eq.(\ref{eq13}) and we obtain the Green's function of slave spin as follows\cite{Nandkishore}
\begin{equation}
G_{\varphi}(k,\omega)\sim\frac{1}{\Delta^{2}+c^{2}k^{2}+\omega^{2}}.\label{eq16}
\end{equation}
The corresponding spectral function can also be easily derived with the form
\begin{equation}
A_{\varphi}(k,\omega+i0^{+})=\delta(\omega^{2}-(\Delta^{2}+k^{2})).\label{eq17}
\end{equation}
Clearly, the spectral function of the slave spin has an excitation gap $\Delta$. Therefore, the physical electron will also acquire a gap with $Z=0$. According to the definition of the orthogonal metal in the paper of Nandkishore, Metlitski and Senthil,\cite{Nandkishore} if a state has a gap for single-particle excitation and the same thermodynamics and transport properties as Landau Fermi liquid, it could be identified as an orthogonal metal. In our case, the physical $c$ electron has an excitation gap while the $f$ fermions form Dirac semimetal. More importantly, the $f$ fermions carry both charge and spin degrees of freedom of the physical $c$ electrons, thus $f$ fermions will contribute to the thermodynamics, charge and spin transports exactly in the same way as real electrons. (The contribution of slave spins can be neglected in the low energy limit, since they are gapped in the disordered state.) Therefore, the $c$ electrons are in a gapped semimetal which is named as orthogonal Dirac semimetal.

Evidently, the orthogonal Dirac semimetal will have the same thermodynamics and transport properties as the usual Dirac semimetal, but it is noted the orthogonal Dirac semimetal is indeed a $Z_{2}$ fractionalized state with the $Z_{2}$ gauge field gapped, thus the only active actor is the slave-fermion which can be defined as a real fractionalized excitation while the usual Dirac semimetal is a confined state of $Z_{2}$ gauge field (see an argument of this point in Appendix C).

Another interesting point is that, in contrast to the usual Dirac semimetal, the $Z_{2}$ fractionalized orthogonal Dirac semimetal should have the subtle order, namely, the topological order \cite{Wen1991,Senthil2001} as argued in Ref. [\onlinecite{Ruegg2012}]. What is needed to point out is however that there is a difference between our case and the one discussed in Ref. [\onlinecite{Ruegg2012}] since the $Z_{2}$ fractionalized orthogonal Dirac semimetal has gapless slave-fermion excitations while in Ref. [\onlinecite{Ruegg2012}], all low-lying excitations for the fractionalized quantum spin Hall insulator are gapped. But this distinction seems irrelevant to the issue of the topological order and we note our case is rather similar to the gapless $Z_{2}$ spin liquid \cite{noteWen2004} in Ref. [\onlinecite{Wen2004}] and the nodal liquid in Ref. [\onlinecite{Senthil2000}]. It is also noted that the authors of Ref. [\onlinecite{Senthil2001}] argued the effect of matter fields to the pure $Z_{2}$ gauge theory and found the topological degeneracy still survives even the matter field has linearly gapless spectrum (the f-fermions are in just this case). The existence of the nontrivial topological order means when considering the system defined in a cylinder (torus), one may find a twofold (fourfold) topological degeneracy of the ground state and a topological term $S_{topo}=-\ln2$ in the entanglement entropy.\cite{Kitaev,Levin,Zhang2011} We expect such nontrivial topological entanglement entropy could be considered as a strong signal to imply the existence of orthogonal Dirac semimetal in future numerical simulations if one finds a gapped metallic state with no symmetry breaking. However, we should emphasize that unlike most of slave-particle approaches, here the fractionalization is irrelevant to the spin-charge separation and it seems this feature is general for any fractionalized states obtained in $Z_{2}$ slave-spin approach.

\section{quantum phase transition of orthogonal Dirac semimetal} \label{sec4}
Having analyzed the properties of the mean-field theory, in this section, we proceed to discuss the stability of the mean-field treatment, which is crucial for the above study of critical behaviors and scaling properties since effect of fluctuation may be fundamentally important near QCP. In certain sense, even one finds mean-field decoupling is unstable, it does not imply the disappearance of orthogonal Dirac semimetal but only shows the critical behavior and phase transition cannot be described by the simple $\varphi^{4}$ theory.

\subsection{Stability of the mean-field decoupling}
In this subsection, we will study the stability of the mean-field treatment which decouples the interacting Hamiltonian Eq. (\ref{eq3}) into two independent Hamiltonians (see, Eqs.(\ref{eq10}) and (\ref{eq11})) with extra self-consistent equations.
Since the stability of mean-field theory in the $Z_{2}$ slave-spin formalism has been discussed in Ref. [\onlinecite{Nandkishore}] with usual Fermi liquid being the normal state (slave-spin condensates), we will follow their treatment but in our case, a crucial difference appears as a Dirac semimetal but not the usual Fermi liquid serves as the normal state in the honeycomb lattice at half-filling. As a matter of fact, this difference leads to an interesting feature that the vanishing density of states of the Dirac semimetal suppresses the
effect of Landau damping even in the conditions without long-range Coulomb interaction, thus the mean-field decoupling is more stable than the one in Ref. [\onlinecite{Nandkishore}] and as a result, the critical Fermi surface (Dirac point) is obtained rather naturally in the honeycomb lattice, in contrast to the case of Ref. [\onlinecite{Nandkishore}].

We now show how one can obtain above statement by following Ref. [\onlinecite{Nandkishore}]. Obviously, the fluctuation above the mean-field theory results from the coupling term between the slave-spin and the slave-fermion in Eq. (\ref{eq3})
\begin{eqnarray}
H_{int}=-t\sum_{\langle ij\rangle\sigma}(\tau_{i}^{x}\tau_{j}^{x}f_{i\sigma}^{\dag}f_{j\sigma}+h.c.)\label{eq18}
\end{eqnarray}
or equivalently one may also use its path integral formalism
\begin{eqnarray}
S_{int}=\int d\tau\left(-t\sum_{\langle ij\rangle\sigma}(\varphi_{i}\varphi_{j}\bar{f}_{i\sigma}f_{j\sigma}+c.c.)\right).\label{eq19}
\end{eqnarray}
Since we are only interested in the physics in low-energy limit, it is helpful to derive an effective theory as
\begin{eqnarray}
&&S=\int d\tau d^{d}x[L_{\varphi}+L_{f}+L_{int}]\\\label{eq20}
&&L_{\varphi}=(\partial_{\tau}\varphi)^{2}+c^{2}(\nabla\varphi)^{2}+r\varphi^{2}+u\varphi^{4}\\\label{eq21}
&&L_{f}=\sum_{\sigma}\bar{\psi}_{\sigma}\gamma_{\mu}\partial_{\mu}\psi_{\sigma}\\\label{eq22}
&&L_{int}=\lambda\sum_{\sigma}(\varphi^{2}\bar{\psi}_{\sigma}W\psi_{\sigma}+c.c.),\label{eq23}
\end{eqnarray}
where we have defined a $4\times4$ matrix $W=\gamma_{2}+i\gamma_{0}\gamma_{3}$ and $L_{\varphi}$, $L_{f}$ come from the free action in Eqs. (\ref{eq13}) and (\ref{eq14}), respectively. And $\lambda$ denotes an effective coupling parameter and can be estimated as $\lambda\sim ta_{0}$ with $a_{0}$ the lattice constant.

It should be emphasized that when the mean-field treatment is applicable (In the mean-field theory, one just neglects the interacting term $L_{int}$.), the quantum critical behaviors can only result from the slave spins since the corresponding quantum Ising model has a definite quantum phase transition while the free Dirac fermions contribute no singularity. Therefore, it is more interesting to study the effect of fluctuations (which comes from the interacting term $L_{int}$ Eq.[\ref{eq23}]) to the slave spins and a one-loop calculation gives the following correction term to action of the slave spins as
\begin{equation}
\lambda^{2}\int d^{d+1}q \frac{N}{4}\sqrt{q^{2}}(1+\frac{q_{x}^{2}+q_{y}^{2}}{2q^{2}})|O(q)|^{2}, \label{eq24}
\end{equation}
where $O(x)=\varphi(x)^{2}$, $q=(\vec{q},\omega)$ and $N=2$ for spin degeneracy with $c\equiv1$. Then, following Ref. [\onlinecite{Nandkishore}], a renormalization-group (RG) argument can be applied in the above action which states the scaling dimension of $\lambda^{2}$ is $\text{dim}[\lambda^{2}]=\frac{2}{\nu}-1-D$ with $D=d+1$, $\nu$ being the critical exponent in the correlation length ($\xi\sim|g|^{-\nu}$ with $g\sim(U-U_{c})/t$ and $U_{c}$ denoting the critical strength of the onsite Coulomb energy at which a QCP locates) of the classical Ising models in D. (The critical exponents of the quantum Ising model in the space dimension of d is identical to the classical Ising model living in d+1 spatial dimensions and this implies the dynamical critical exponent ($z$) is equal to one.)

According to the scaling dimension of $\lambda^{2}$, the correction term given by Eq. (\ref{eq24}) to slave spins, which results from integrating over Dirac fermions to lowest order, is irrelevant when $\nu>\frac{2}{D+1}$ is satisfied. For our case, we are considering a two-dimensional honeycomb lattice thus our system is living in d=2 and one should use $\nu=0.63$, a critical exponent for classical three-dimensional Ising model. Therefore, it is clear that the induced correction term for slave spins is irrelevant in the sense of RG, which means the coupling term between the slave-spin and the slave-fermion cannot change the low-energy physics obtained from the simple mean-field approximation. Moreover, the stability of the mean-field treatment suggests that quantum critical behaviors are controlled only by the quantum Ising model of slave-spins and all critical exponents fall in the Ising universal class.

However, our result is quite different from the case of orthogonal metal \cite{Nandkishore} where generically the mean-field decoupling is only valid in the presence of the long-range Coulomb interaction. In our system no such long-range interaction is needed. The main reason is that if the slave-fermions $f$ form a usual Fermi liquid, they will always contribute a standard Landau damping term with an extra constant being the density of state (DOS) at the Fermi surface for slave-spins. This nonzero DOS term is not irrelevant unless one adds the long-range Coulomb interaction. In contrast, if the Dirac semimetal is considered, as in our case, slave spins will not gain a nonzero DOS term but a term given by Eq. (\ref{eq24}), just like the long-range Coulomb interaction in two space dimensions since for the free Dirac fermions, its DOS at the Fermi surface (at the Dirac points in our case) vanishes (In d=2, the Coulomb interaction is $V(\vec{q})\sim\frac{e^{2}}{|\vec{q}|}$ with $V(\vec{x})\sim\frac{e^{2}}{|\vec{x}|}$.\cite{Herbut2006}). Moreover, since no Fermi surface (only separated Dirac points) appears in our case, we expect that no usual Landau damping appears from the long-wavelength deformations of
the `Fermi surface' due to disappearing of continuous rotation symmetry in lattice models.\cite{Nandkishore}

After all, the mean-field treatment in the last section is stable to fluctuation resulting from coupling between the slave-spin and the slave-fermion even without introducing the long-range Coulomb interaction. Then we will proceed to use the mean-field approximation to analyze the critical behaviors and corresponding scaling properties near QCP.

\subsection{Critical behaviors and scaling properties near quantum phase transition of orthogonal Dirac semimetal}
In this subsection, we will study the critical behaviors near the quantum phase transition from orthogonal Dirac semimetal to usual Dirac semimetal by using the effective action Eqs. (\ref{eq20}) and (\ref{eq21}) which reflects the decoupling in the mean-field treatment.

Obviously, the usual Dirac semimetal is well-understood and its low-energy excitation is described by the Dirac quasiparticle according to the discussion in the Sec. \ref{sec3} when approaching the QCP
\begin{equation}
G(k)=Z\frac{i\gamma_{\mu}k_{\mu}}{k^{2}}, \label{eq25}
\end{equation}
where the quasiparticle spectral weight $Z=\langle\tau^{x}\rangle^{2}\sim g^{2\beta}$. (One can approach QCP from the orthogonal Dirac semimetal with vanished excitation gap as well.) Besides, based on a scaling argument in Ref. [\onlinecite{Herbut2009}], the quasiparticle spectral weight for $z=1,d=2$ can be written as $Z\sim g^{v\eta_{c}}$, where an anomalous dimension $\eta_{c}$ for physical electrons is defined. Thus, a crucial result in the present paper is obtain as $\eta_{c}=1+\eta$. (In fact, for $d=2,z=1$, this result still holds even if the decoupling approach is invalid. To derive above result, we have used two scaling relations $\alpha+2\beta+\gamma=2$,$\nu(2-\eta)=\gamma$ and the hyperscaling law $2-\alpha=\nu(d+z)$ with $\alpha,\beta,\gamma,\nu,\eta$ being the critical exponents of the quantum Ising model.) This means that at QCP the Dirac quasiparticle is destroyed completely ($Z=0$) since a large anomalous dimension appears ($\eta_{c}\simeq1.036,\eta\simeq0.036$) and the local DOS will behave as $N(\omega)\sim|\omega|^{1+\eta_{c}}$ which is rather different from its counterpart in Dirac semimetal ($N(\omega)\sim|\omega|$). In the sense of the critical Fermi surface of Senthil,\cite{Sethil2008} the six Dirac points are indeed critical at QCP with nontrivial power-law behaviors in the spectral function (or DOS) of physical electrons. Thus, we may call them critical Dirac points following the similar spirit of the critical Fermi surface. However, to our surprised, the quasiparticle picture does not break down at QCP because slave spins and slave fermions can be considered as the real quasiparticle at criticality, respectively. Therefore, quantum critical behaviors are readily to be obtained in terms of these quasiparticles. For example, at QCP the specific heat is $C_{v}\sim T^{2}$ which contributes from both slave spins and slave fermions. And the electrical conductivity at QCP is easily obtained as  $\sigma(\omega)=\frac{\pi}{4}\frac{e^{2}}{2\pi}$ which is identical to the one in the usual Dirac semimetal since slave fermions carry all quantum number of physical electrons and they form Dirac semimetal like physical quasiparticles in the Dirac semimetal phase.

\section{Extensions and possible relations to other models and approaches} \label{sec5}
In this section, we will briefly argue the instability of the orthogonal Dirac semimetal to other phases and discussion possible relations of our results obtained in previous sections to other models and approaches. A careful reader may wonder whether one can construct some exactly soluble models for the orthogonal Dirac semimetal. In fact, it is straightforward to construct possible exact soluble models by following the idea of Ref. [\onlinecite{Nandkishore}]. The only difference needed to be careful is to put their model from square lattice to the honeycomb lattice. Here we focus on the special features of the honeycomb lattice and do not discuss it in detail.

\subsection{Instability of the orthogonal Dirac semimetal to sublattice pairing states and fractionalized quantum spin Hall insulator}
In principle, the orthogonal Dirac semimetal proposed may be seen as a "Fermi liquid" which can suffer from further transitions to other fractionalized states. For example, if a next-nearest-neighbor hopping term is added into the Hubbard model, the particle-hole symmetry will be violated, thus the terms involving the Lagrange multiplier $\lambda_{i}$ cannot be ignored, which may lead to a paring instability of slave fermions $f_{\sigma}$. Particularly, when a pairing is realized between the next-nearest-neighbor sites, the resulting state is like the so-called sublattice pairing states (SPS),\cite{Wang2010,Lu2011,Clark} which has a gap both for charge and spin excitations and has been argued to be a promising candidate for gapped spin liquid state found in the numerical simulation. However, detail treatment of this sublattice pairing states is beyond our present paper and it is interesting to attack this issue in future work.

In addition, one may consider a spin-orbit coupling between next-nearest-neighbor sites
\begin{equation}
H_{SO}=i\lambda\Sigma_{\langle\langle ij\rangle\rangle}\Sigma_{\sigma\sigma'}\nu_{ij}\sigma^{z}_{\sigma\sigma'}c^{\dag}_{i\sigma}c_{j\sigma'},\label{eq26}
\end{equation}
where $\lambda$ is the spin-orbit coupling and $\nu_{ij}$ is defined as in Ref. [\onlinecite{Rachel}] with $\langle\langle \cdot \rangle\rangle$ denoting next-nearest-neighbor sites. Therefore, the resulting model will be the celebrated Kane-Mele-Hubbard model which is a good starting point for studying interaction effect on two-dimensional
topological insulators.\cite{Rachel} Obviously, due to this spin-orbit coupling, the usual Dirac semimetal will smoothly evolve into the two-dimensional
topological insulator (or quantum spin Hall state) while the exotic orthogonal Dirac semimetal will inevitably open an excitation gap at Dirac points, which will give rise to the exotic fractionalized quantum spin Hall insulator found by R\"{u}egg and Fiete.\cite{Ruegg2012} This phase could be dubbed an ``orthogonal spin Hall state" or an ``orthogonal helical metal" according to one's taste.\cite{Nandkishore} Before ending this subsection, we should emphasize that a
gapped spin liquid has also been found in Kane-Mele-Hubbard model when spin-orbit coupling $\lambda$ is small.\cite{Hohenadler2011,Zheng2011,Yu2011} However, our current approach cannot capture this subtle issue and we expect future work may clarify the nature of this gapped spin liquid and its possible relation to the orthogonal spin Hall state.

\subsection{Comparison the $Z_{2}$ fractionalization approach to the Gross-Neveu theory}
Beside the above possible transitions, it is also noted that our results are rather different from the treatment of Herbut who did not use fractionalization but started with usual order parameters, e.g. spin/charge density wave (SDW/CDW), and then coupled them minimally to Dirac fermions.\cite{Herbut2006,Herbut} The resulting theory is described by the Gross-Neveu theory with a large anomalous dimension for the order parameter and is argued to be outside the Ginzburg-Landau-Wilson paradigm.
But, in the effective Gross-Neveu theory, the physical electrons only have a small anomalous dimension while our treatment gives a rather large anomalous dimension for physical electrons since they are fractionalized at QCP and a large anomalous dimension is naturally expected. Therefore, the large anomalous dimension of physical electrons and the related pow-law behavior in the local DOS may be considered as a smoke-gun prediction of our $Z_{2}$ slave-spin theory for Hubbard model in the honeycomb lattice when comparing to the usual effective Gross-Neveu theory.

\subsection{Relations to the fermionic dual approach of Mott transition and the gauge/gravity duality}
Moreover, it is also interesting to see whether orthogonal Dirac semimetal and the corresponding quantum phase transition can be described by extending the dual approach developed in Mott transition from Fermi liquid to quantum spin liquid.\cite{Mross,Zhong} However, we have tried and found that such a description is not feasible due to lack of Fermi surface when only half-filling is involved. In contrast, a dual description identical to Refs. [\onlinecite{Mross}, \onlinecite{Zhong}] can be constructed when deviating from half-filling since in this case the chemical potential is nonzero and a Fermi surface instead of Dirac points is formed but this is not interesting for our present paper and we will not discuss it further.

The last relation we considered is the link to the gauge/gravity duality \cite{Sachdev2012,Hartnoll,McGreevy} from string theories. Based on some arguments in holographic entanglement entropy,\cite{Ogawa,Huijse} which is an remarkable application of the gauge/gravity duality to calculate entanglement entropy of certain dual strongly correlated field theories, only gapless metallic states with fermionic dynamical critical exponent large than one can be realized in the current models in the gauge/gravity duality for non-Fermi liquids if a Fermi surface is assumed. Thus, it is clear that the exotic orthogonal metal (the simplest non-Fermi liquid) cannot be found in these models since it is gapped and its fermionic dynamical critical exponent is equal to one. For the case of orthogonal Dirac semimetal, to our knowledge, no such state has been discovered in existing gauge/gravity duality. Therefore, we suspect that the nature of exotic orthogonal metal, orthogonal Dirac semimetal and their corresponding quantum criticality may not be captured by existing models in gauge/gravity duality and more study is desired towards this direction.

\section{Discussion and Conclusion}\label{sec6}
In the present paper we have shown that a $Z_{2}$ fractionalized metallic state called orthogonal Dirac semimetal can exist in the $Z_{2}$ slave-spin representation of Hubbard model in the honeycomb lattice at half-filling. The orthogonal Dirac semimetal can survive when slave spins become disordered. This state has the same thermodynamics and transport as usual Dirac semimetal but with gapped singe-partice excitation and nontrivial topological order. Moreover, the quantum phase transition (QPT) from Dirac semimetal to fractionalized orthogonal Dirac semimetal is analyzed by the mean-field decoupling and its criticality can be described by the universality class of 2+1D Ising model. The result that the physical electron gains a large anomalous dimension at QCP presents the fingerprint of our slave-spin theory. It should be emphasized that in contrast to most of slave-particle approaches, the $Z_{2}$ fractionalization involved in the $Z_{2}$ slave-spin representation is irrelevant to the spin-charge separation. Thus, the orthogonal Dirac semimetal should be considered as a simple non-Fermi liquid but without spin-charge separation. It seems this feature is general for any fractionalized states obtained in $Z_{2}$ slave-spin approach.

In addition, we have also constructed a path integral formalism for the $Z_{2}$ slave-spin representation of Hubbard model and discussed possible relations to effective Gross-Neveu theory and gauge/gravity duality. Furthermore, the instability of orthogonal Dirac semimetal and the possible relation to the SPS spin liquid and the fractionalized quantum spin Hall insulator are briefly analyzed as well. In our opinion, the $Z_{2}$ slave-spin representation and its mean-field results seem to work well when the onsite $U$ is not too large, otherwise, slave boson or slave rotor approach may be more economic to capture physics of quantum spin liquids. Besides, we expect the exotic orthogonal Dirac semimetal may be realized in the sophisticated experiments of ultracold atoms in the honeycomb optical lattices.\cite{Bloch,Goldman,Bermudez} It will be also interesting to see whether topological insulator in 3D may provides a realization of this state if strong coupling is accomplished carefully.\cite{Hasan2010,Qi2011} Moreover, there may also exist an interesting extension of the present analysis of orthogonal Dirac semimetal to the bilayer graphene with a non-vanishing density of state.\cite{Nandkishore2010,Nandkishore2012} We hope our findings may be helpful for future studies in $Z_{2}$ slave-spin theory and non-Fermi liquid phases in honeycomb lattice.

\begin{acknowledgments}
The authors would like to thank Y. F. Wang for useful discussion. We also thank F. Mezzacapo, A. R\"{u}egg
and M. A. Martin-Delgado for helpful suggestions. The work was supported partly by NSFC, the Program for NCET, the Fundamental Research Funds for the Central Universities and the national program for basic research of China.
\end{acknowledgments}

\appendix

\section{The explanation of why the slave-spin does not carry charge of the physical electron}
Here we should emphasize that although the physical $c$ electron has been fractionalized into
an auxiliary fermion $f_{\sigma}$ and a slave spin $\tau^{x}$, the quantum number of the electron (the spin-$\frac{1}{2}$ and the charge $e$) are both carried by the $f$ fermion, which is quite different from slave boson or slave rotor approaches where the charge and spin degree of freedom are solely carried by bosonic particles and fermionic spinons, respectively. This point is not noticed until the recent interesting paper \cite{Nandkishore} but has crucial influence on the correct interpretation of the disordered state of the slave spin. As argued in Refs. [\onlinecite{Nandkishore}] , a U(1) rotation of physical electron $d$ can only be matched by a U(1) rotation of $f$ fermion while the slave spin $\tau^{x}$ do not change because it is purely real. Therefore electric charge must be only carried by $f$ fermion but not the slave spin since it corresponds to the Noether charge of the U(1) symmetry.

\section{path integral for the quantum Ising model in transverse field}
The quantum Ising model in transverse field is defied as\cite{Sachdev2011}
\begin{equation}
\hat{H}_{I}=-J\sum_{\langle ij\rangle\sigma}(\tau_{i}^{z}\tau_{j}^{z}+h.c.)-K\sum_{i}\tau_{i}^{x}
\end{equation}
where a ferromagnetic coupling $J>0$ is assumed and $K$ represents the the transverse external field.

At first glance, one may directly use the coherent state of spin operators in constructing the path integral representation, (One can find a brief but useful introduction to this issue in Ref. [\onlinecite{Sachdev2011}]) however, this will lead to an extra topological Berry phase term and is not easy to utilize practically. An alterative approach is to use the eigenstates of spin operator $\tau^{x}$ or $\tau^{z}$ as the basis for calculation.\cite{Stratt}
One will see this approach is free of the topological Berry phase term and give rise to a rather simple formalism.  Therefore, to construct a useful path integral representation, we will follow Ref. [\onlinecite{Stratt}].

First of all, we consider the orthor-normal basis of $N_{s}$-Ising spins as
\begin{equation}
|\sigma\rangle\equiv|\sigma_{1}\rangle|\sigma_{2}\rangle|\sigma_{2}\rangle\cdot\cdot\cdot|\sigma_{N}\rangle
\end{equation}
with $\sigma_{i}=\pm1$ and define
\begin{equation}
\tau_{i}^{z}|\sigma\rangle=\sigma_{i}|\sigma\rangle,
\end{equation}
\begin{equation}
\tau_{i}^{x}|\sigma\rangle
=|\sigma_{1}\rangle|\sigma_{2}\rangle|\sigma_{3}\rangle\cdot\cdot\cdot|-\sigma_{i}\rangle\cdot\cdot\cdot|\sigma_{N}\rangle.
\end{equation}
Then the partition function $Z=Tr(e^{-\beta \hat{H}})$ can be represented as
\begin{eqnarray}
Z=\sum_{\{\sigma\}=\pm1}\prod_{n=1}^{N}e^{\epsilon J\sum_{\langle ij\rangle}\sigma_{i}(n)\sigma_{j}(n)}\langle\sigma(n+1)|e^{\epsilon K\sum_{i}\tau_{i}^{x}}|\sigma(n)\rangle \nonumber
\end{eqnarray}
where $\epsilon$N=$\beta$.
The calculation of $\langle\sigma(n+1)|e^{\epsilon K\sum_{i}\tau_{i}^{x}}|\sigma(n)\rangle$ is straightforward
by exponentiating the $\tau_{i}^{x}$ matrix and one gets
\begin{eqnarray}
\langle\sigma(n+1)|e^{\epsilon K\sum_{i}\tau_{i}^{x}}|\sigma(n)\rangle&&=\frac{1}{2}(e^{\epsilon K}+e^{-\epsilon K}\sigma_{i}(n)\sigma_{i}(n+1)),\nonumber \\
&&=e^{a\sigma_{i}(n)\sigma_{i}(n+1)+b}
\end{eqnarray}
where $a=\frac{1}{2}[\ln\cosh(\epsilon K)-\ln\sinh(\epsilon K)]$ and $b=\frac{1}{2}[\ln\cosh(\epsilon K)+\ln\sinh(\epsilon K)]$.
Therefore, the resulting path integral formalism for the quantum Ising model in transverse field is
\begin{equation}
Z=\sum_{\{\sigma\}=\pm1}\prod_{n=1}^{N}e^{\epsilon J\sum_{\langle ij\rangle}\sigma_{i}(n)\sigma_{j}(n)+\sum_{i}a\sigma_{i}(n)\sigma_{i}(n+1)+N_{s}b}.
\end{equation}

Further, if one assumes the model is defined in a hyper-cubic lattice in space dimension of d, an effective theory can be derived as
\begin{equation}
Z=\int D\phi \delta(\phi^{2}-1) e^{-\int d\tau d^{d}x \frac{1}{2g}[(\partial_{\tau}\phi)^{2}+c^{2}(\nabla\phi)^{2}]},
\end{equation}
where $\frac{1}{2g}=(\frac{a\epsilon}{a_{0}^{d}})^{\frac{d+1}{2}}$ with $a_{0}$ being the lattice constant and $c^{2}=\frac{Ja_{0}^{d-2}}{a\epsilon}$. Moreover, in the effective theory, $\phi$ corresponds to $\tau^{z}$ while $\tau^{x}$ gives the kinetic energy term in imaginary time. Then, the standard $\phi^{4}$ theory is obtained by relaxing the hard constraint $\phi^{2}=1$ while introducing a potential energy term,
\begin{equation}
Z=\int D\phi e^{-\int d\tau d^{d}x [(\partial_{\tau}\phi)^{2}+c^{2}(\nabla\phi)^{2}+r\phi^{2}+u\phi^{4}]},
\end{equation}
where $r,u$ are effective parameters depending on microscopic details.

\section{$Z_{2}$ gauge theory formalism for $Z_{2}$ spin-slave representation}
Here, we derive the $Z_{2}$ gauge theory formalism for $Z_{2}$ spin-slave representation by using Eqs. (\ref{eq8}) and (\ref{eq9}).

First, it is noted in Eq. (\ref{eq9}) the constraint can be dropped because we are only interested in non-magnetic solutions. Then using the familiar Hubbard-Stratonovich transformation to decouple the coupling term between the slave-spin and the slave-fermion in Eq. (\ref{eq9}), one obtains
\begin{eqnarray}
S&&=\int d\tau\sum_{i}[\bar{f}_{i\sigma}\partial_{\tau}f_{i\sigma}+\tilde{a}(\partial_{\tau}\varphi_{i})^{2}] \nonumber\\
&&-\int d\tau\sum_{\langle ij\rangle}(J_{ij}\varphi_{i}\varphi_{j}+\tilde{t}_{ij}\bar{f}_{i\sigma}f_{j\sigma}+c.c.)\nonumber\\
&&+\int d\tau\frac{1}{t}\sum_{\langle ij\rangle}J_{ij}\tilde{t}_{ij}
\end{eqnarray}
where the summation in the imaginary time is transformed to an integral and $J_{ij}$, $\tilde{t}_{ij}$ are the auxiliary field introduced in the Hubbard-Stratonovich transformation, respectively. The usual mean-field approximation can be recovered by treating $J_{ij},\tilde{t}_{ij}$ as static variable. Beyond mean-field approximation, one may allow a phase fluctuation in the static $J_{ij},\tilde{t}_{ij}$, which are labeled as $J_{ij}^{MF},\tilde{t}_{ij}^{MF}$ to emphasize they are solutions of the mean-field approximation. However, since the original Hamiltonian Eq. (\ref{eq3}) only has the local $Z_{2}$ symmetry, the phase fluctuation is nothing but the expected $Z_{2}$ gauge field defined in the link between two sites. \cite{Ruegg2012} Therefore, we have
\begin{equation}
J_{ij}\simeq J_{ij}^{MF}\sigma_{ij},\tilde{t}_{ij}\simeq\tilde{t}_{ij}^{MF}\sigma_{ij}
\end{equation}
and
\begin{eqnarray}
S_{eff}&&=\int d\tau\sum_{i}[\bar{f}_{i\sigma}\partial_{\tau}f_{i\sigma}+\tilde{a}(\partial_{\tau}\varphi_{i})^{2}] \nonumber\\
&&-\int d\tau\sum_{\langle ij\rangle}(J_{ij}^{MF}\sigma_{ij}\varphi_{i}\varphi_{j}+\tilde{t}_{ij}^{MF}\sigma_{ij}\bar{f}_{i\sigma}f_{j\sigma}+c.c.)\nonumber
\end{eqnarray}
where $\sigma_{ij}$ denotes the dynamical $Z_{2}$ gauge field and terms without dynamics are neglected. Much physics can be seen from the above action. If the slave-spin ($\varphi$) condensates, the $Z_{2}$ gauge field will be confined by the Higgs mechanism while the slave-spin and slave-fermion have to bind into physical electrons as the only low-energy excitation. In contrast, the disordered slave-spin ($\varphi$) is gapped and can be integrated out to generate an effective ``kinetic energy" term for the $Z_{2}$ gauge field. \cite{Ruegg2012} In principle, the $Z_{2}$ gauge field can be in the deconfined state in 2+1D if the effective kinetic energy is large enough to suppress the fluctuation of the $Z_{2}$ gauge field. Then, in the deconfined state, one can treat the gauge field as a static background (The excitation of the gauge field, the $Z_{2}$-vortex (vison) is gapped.), where the slave-fermion is almost free and can be defined as real excitation, thus one has a $Z_{2}$ fractionalized state indeed.

\section{A self-consistent mean-field solution}
In this Appendix, we give the single-site self-consistent mean-field solution for Eqs. [\ref{eq10}] and [\ref{eq11}] without the Lagrange multiplier.

Following the treatment of Sec.4 in Ref.[\onlinecite{Ruegg}] and using Eqs. [\ref{eq10}] and [\ref{eq11}], the self-consistent equation for $g\equiv \langle\tau^{x}\rangle^{2}$ is obtained as
\begin{equation}
1=\frac{4\bar{\varepsilon}}{\sqrt{(U/4)^{2}+16\bar{\varepsilon}^{2}g}}
\end{equation}
where we have defined $\bar{\varepsilon}=\int_{0}^{\infty}d\varepsilon\rho(\varepsilon)\varepsilon$ with $\rho(\varepsilon)$ denoting density of states of free f-fermions.  When $g=0$, one finds the critical coupling $U_{c}=16\bar{\varepsilon}\simeq 12.444t$. For $U>U_{c}$, the slave-spin does not form ferromagnetic order while the ferromagnetic order appears with nonzero $g=1-(U/U_{c})^{2}$ for $U<U_{c}$.
As for the f-fermions, they can be described by the following Hamiltonian
\begin{equation}
H=-t g\sum_{\langle ij\rangle\sigma}(f_{i\sigma}^{\dag}f_{j\sigma}+H.c.).
\end{equation}


\begin{thebibliography}{99}

\bibitem{Sachdev2011}
S. Sachdev, \textit{Quantum Phase Transition}, 2nd ed. (Cambridge University Press, Cambridge, England, 2011).

\bibitem{Sachdev2003}
S. Sachdev, Rev. Mod. Phys \textbf{75}, 913 (2003).

\bibitem{Rosch}
H. V. L\"{o}hneysen, A. Rosch, M. Vojta and P. W\"{o}lfle, Rev. Mod. Phys \textbf{79}, 1015 (2007).

\bibitem{Sachdev2008}
S. Sachdev, Nature Phys \textbf{4}, 173 (2008).

\bibitem{Powell}
B. J. Powell and R. H. McKenzie, Rep. Prog. Phys. \textbf{74} 056501 (2011).

\bibitem{Georges1996}
A. Georges, G. Kotliar, W. Krauth and M. J. Rozenberg, Rev. Mod. Phys. \textbf{68}, 13 (1996).

\bibitem{Schollwock}
U. Schollw\"{o}ck, Rev. Mod. Phys. \textbf{77}, 259 (2005).

\bibitem{Wen2004}
Xiao-Gang Wen, \textit{Quantum Field Theory of Many-Body Systems}, (Oxford Graduate Texts, New York, 2004).

\bibitem{Wen}
P. A. Lee, N. Nagaosa and X. G. Wen, Rev. Mod. Phys \textbf{78}, 17 (2006).

\bibitem{Kitaev}
A. Kitaev and J. Preskill, Phys. Rev. Lett. \textbf{96}, 110404 (2006).

\bibitem{Levin}
M. Levin and X.-G. Wen, Phys. Rev. Lett. \textbf{96}, 110405 (2006).

\bibitem{Metzner}
W. Metzner, M. Salmhofer, C. Honerkamp, V. Meden and K. Sch\"{o}nhammer,  Rev. Mod. Phys \textbf{84}, 299 (2012).

\bibitem{Senthil2003}
T. Senthil, S. Sachdev and M. Vojta, Phys. Rev. Lett. \textbf{90}, 216403 (2003).

\bibitem{Senthil2004}
T. Senthil, M. Vojta and S. Sachdev, Phys. Rev. B \textbf{69}, 035111 (2004).

\bibitem{Senthil3}
T. Senthil et al., Science \textbf{303}, 1490 (2004).

\bibitem{Senthil4}
T. Senthil, L. Balents, S. Sachdev, A. Vishwanath and M. P. A. Fisher, Phys. Rev. B \textbf{70}, 144407 (2004).

\bibitem{Florens}
S. Florens and A. Georges, Phys. Rev. B \textbf{70}, 035114 (2004).

\bibitem{Lee2005}
S. S. Lee and P. A. Lee, Phys. Rev. Lett. \textbf{95}, 036403 (2005).

\bibitem{Pepin2005}
C. P\'epin, Phys. Rev. Lett. \textbf{94}, 066402 (2005).

\bibitem{Kim2006}
K. S. Kim, Phys. Rev. Lett. \textbf{97}, 136402 (2006).

\bibitem{Senthil2008}
T. Senthil, Phys. Rev. B \textbf{78}, 045109 (2008).

\bibitem{Kim2010}
K. S. Kim and C. L. Jia, Phys. Rev. Lett. \textbf{104}, 156403 (2010).

\bibitem{Senthil2010}
T. Grover and T. Senthil, Phys. Rev. B \textbf{81}, 205102 (2010).

\bibitem{Meng}
Z. Y. Meng, T. C. Lang, S. Wessel, F. F. Assaad, and A. Muramatsu, Nature (London) \textbf{464}, 847 (2010).

\bibitem{Sorella1992}
S. Sorella and E. Tosatti, EPL \textbf{19}, 699 (1992).

\bibitem{Herbut2006}
I. F. Herbut, Phys. Rev. Lett. \textbf{97}, 146401 (2006).

\bibitem{Kotov2010}
V. N. Kotov, B. Uchoa, V. M. Pereira, F. Guinea and A. H. C. Neto, Rev. Mod. Phys \textbf{84}, 1067 (2012).

\bibitem{Yan}
S.-M. Yan, D. A. Huse and S. R. White,  Science \textbf{332}, 1173 (2011).

\bibitem{Sorella2012}
Y. O. Sandro Sorella and S. Yunoki, arXiv:cond-mat/1207.1783v1 (2012).

\bibitem{Wang2010}
F. Wang, Phys. Rev. B \textbf{82}, 024419 (2010).

\bibitem{Tran2011}
M.-T. Tran and K.-S. Kim, Phys. Rev. B \textbf{83}, 125416 (2011).

\bibitem{Lu2011}
Y.-M. Lu and Y. Ran, Phys. Rev. B \textbf{84}, 024420 (2011).

\bibitem{Clark}
B. K. Clark, D. A. Abanin and S. L. Sondhi, Phys. Rev. Lett. \textbf{107}, 087204 (2011).

\bibitem{Mezzacapo2012}
F. Mezzacapo and M. Boninsegni, Phys. Rev. B \textbf{85}, 060402(R) (2012).

\bibitem{Mezzacapor}
We thank F. Mezzacapo for pointing out this point.

\bibitem{Nandkishore}
R. Nandkishore, M. A. Metlitski and T. Senthil, Phys. Rev. B \textbf{86}, 045128 (2012).

\bibitem{deMedici}
L. de'Medici, A. Georges and S. Biermann, Phys. Rev. B \textbf{72}, 205124 (2005).

\bibitem{Hassan}
S. R. Hassan and L. de Medici, Phys. Rev. B \textbf{81}, 035106 (2010).

\bibitem{Ruegg}
A. R\"{u}egg, S. D. Huber and M. Sigrist, Phys. Rev. B \textbf{81}, 155118 (2010).

\bibitem{Yu}
R. Yu and Q. Si, Phys. Rev. B \textbf{84}, 235115 (2011).

\bibitem{Zhong2012}
Y. Zhong, K. Liu, Y. Q. Wang and H.-G. Luo, arXiv:cond-mat/1203.0635 (2012).

\bibitem{Paul}
I. Paul, C. P\'epin and M. R. Norman, Phys. Rev. Lett. \textbf{98} 026402 (2007).

\bibitem{Pepin}
C. P\'epin, Phys. Rev. Lett. \textbf{98} 206401 (2007).

\bibitem{Pepin2008}
C. P\'epin, Phys. Rev. B \textbf{77}, 245129 (2008).

\bibitem{Paul2008}
I. Paul, C. P\'epin and M. R. Norman, Phys. Rev. B \textbf{78} 035109 (2008).

\bibitem{Delgado}
We thank M. A. Martin-Delgado for pointing out this point.

\bibitem{Bloch}
I. Bloch, J. Dalibard and W. Zwerger, Rev. Mod. Phys. \textbf{80}, 885 (2008).

\bibitem{Goldman}
N. Goldman, A. Kubasiak, A. Bermudez, P. Gaspard, M. Lewenstein, and M. A. Martin-Delgado,
Phys. Rev. Lett. \textbf{103}, 035301 (2009).

\bibitem{Bermudez}
A. Bermudez, N. Goldman, A. Kubasiak, M. Lewenstein and M. A. Martin-Delgado,
New J. Phys. \textbf{12}, 033041 (2010).

\bibitem{Neto}
A. H. C. Neto, F. Guinea, N. M. R. Peres, K. S. Novoselov and A. K. Geim, Rev. Mod. Phys. \textbf{81}, 109
(2009).

\bibitem{Hasan2010}
M. Z. Hasan and C. L. Kane, Rev. Mod. Phys. \textbf{82}, 3045 (2010).

\bibitem{Qi2011}
X.-L. Qi and S.-C. Zhang, Rev. Mod. Phys. \textbf{83}, 1057 (2011).

\bibitem{Nandkishore2010}
R. Nandkishore and L. Levitov, Phys. Rev. Lett. \textbf{104}, 156803 (2010).

\bibitem{Nandkishore2012}
R. Nandkishore and L. Levitov, Phys. Scr. \textbf{T146}, 014011 (2012).

\bibitem{Hermele2007}
M. Hermele, Phys.Rev.B \textbf{76}, 035125 (2007).

\bibitem{Ruegg2012}
A. R\"{u}egg and G. A. Fiete, Phys. Rev. Lett. \textbf{108} 046401 (2012).

\bibitem{Senthil2000}
T. Senthil and M. P. A. Fisher, Phys. Rev. B \textbf{62}, 7850 (2000).

\bibitem{Stratt}
R. M. Stratt, Phys. Rev. Lett. \textbf{53}, 1305 (1984).

\bibitem{Rueggpr}
A. R\"{u}egg has also obtained similar formalism (in private communication).

\bibitem{Vaezi2011}
M. Mardani, M.-S. Vaezi and A. Vaezi, arXiv:cond-mat/1111.5980v2 (2011).

\bibitem{Pesin}
D. Pesin and L. Balents, Nature Phys. \textbf{6}, 376 (2010).

\bibitem{Rachel}
S. Rachel and K. Le Hur, Phys. Rev. B \textbf{82}, 075106 (2010).

\bibitem{Wen1991}
X.-G. Wen, Phys. Rev. B \textbf{44}, 2664 (1991).

\bibitem{Senthil2001}
T. Senthil and M. P. A. Fisher, Phys. Rev. B \textbf{63}, 134521 (2001).

\bibitem{noteWen2004}
See section 9.2.5 in Ref. [\onlinecite{Wen2004}] for a brief discussion on a gapless $Z_{2}$ spin liquid.

\bibitem{Zhang2011}
Y. Zhang, T. Grover and A. Vishwanath, Phys. Rev. Lett. \textbf{107}, 067202 (2011).

\bibitem{Herbut}
I. F. Herbut, V. Juricic and O. Vafek, Phys. Rev. B \textbf{80}, 075432 (2009).

\bibitem{Herbut2009}
I. F. Herbut, V. Juricic, and B. Roy , Phys. Rev. B \textbf{79}, 085116 (2009).

\bibitem{Sethil2008}
T. Senthil, Phys. Rev. B \textbf{78}, 035103 (2008).

\bibitem{Hohenadler2011}
M. Hohenadler, T.-C. Lang and F. F. Assaad, Phys. Rev. Lett. \textbf{106} 100403 (2011).

\bibitem{Zheng2011}
D. Zheng, G.-M. Zhang and C.-J Wu, Phys. Rev. B \textbf{84}, 205121 (2011).

\bibitem{Yu2011}
S.-L. Yu, X. C. Xie and J.-X. Li, Phys. Rev. Lett. \textbf{107}, 010401 (2011).

\bibitem{Mross}
D. F. Mross and T. Senthil, Phys. Rev. B \textbf{84}, 165126 (2011).

\bibitem{Zhong}
Y. Zhong, K. Liu and H.-G. Luo, Phys. Rev. B \textbf{85}, 075106 (2012).

\bibitem{Sachdev2012}
S. Sachdev, Annu. Rev. Condens. Matter Phys. \textbf{3}, 5.1 (2012).

\bibitem{Hartnoll}
S. A. Hartnoll,  arXiv:hep-th/0903.3246 (2010).

\bibitem{McGreevy}
J. McGreevy, Adv. High Energy Phys. 2010:723105 (2010).

\bibitem{Ogawa}
N. Ogawa, T. Takayanagi and T. Ugajin, J. High Energy Phys. 01, 125 (2012).

\bibitem{Huijse}
L. Huijse, S. Sachdev and B. Swingle, Phys. Rev. B \textbf{85}, 035121 (2012).

\end{thebibliography}
\end{document}